\documentclass{article}
\usepackage{spconf}
\usepackage{amsmath,pgfplots}
\usepackage{cite}
\input{mysymbol.sty}

\usepackage{amsfonts,amssymb}
\usepackage{hyperref}
\usepackage{mathrsfs}
\usepackage{balance}
\usepackage{mathtools}
\usepackage[boxruled,linesnumbered]{algorithm2e}

\usepackage[mode=buildnew]{standalone}
\usepackage{graphicx}
\usepackage{caption}
\usepackage{subcaption,enumitem}

\def\QED{{\setlength{\fboxsep}{0pt}\setlength{\fboxrule}{0.2pt}\fbox{\rule[0pt]{0pt}{1.3ex}\rule[0pt]{1.3ex}{0pt}}}}

\pgfplotsset{colormap/bluered}

\def\1{{\boldsymbol 1}}

\ninept

\title{An underparametrized Deep Decoder Architecture for Graph Signals}
\name{Samuel Rey$^\dagger$, Antonio G. Marques$^\dagger$, and Santiago Segarra$^\ddagger$
\thanks{This work was supported by the Spanish grants MINECO TEC2016-75361-R, Instituto de Salud Carlos III DTS17/00158, and FPU17/04520.}}
\address{$^\dagger$Dept. of Signal Theory and Communications, King Juan Carlos University, Madrid, Spain\\
    $^\ddagger$Dept. of Electrical and Computer Engineering, Rice University, Houston, TX, USA}

\begin{document}

\maketitle
\begin{abstract}
	While deep \textit{convolutional} architectures have achieved remarkable results in a gamut of supervised applications dealing with images and speech, recent works show that deep \textit{untrained non-convolutional} architectures can also outperform state-of-the-art methods in several tasks such as image compression and denoising. Motivated by the fact that many contemporary datasets have an irregular structure different from a 1D/2D grid, this paper generalizes untrained and underparametrized non-convolutional architectures to signals defined over irregular domains represented by graphs. The proposed architecture consists of a succession of layers, each of them implementing an upsampling operator, a linear feature combination, and a scalar nonlinearity. A novel element is the incorporation of upsampling operators accounting for the structure of the supporting graph, which is achieved by considering a systematic graph coarsening approach based on hierarchical clustering. The numerical results carried out in synthetic and real-world datasets showcase that the reconstruction performance can improve drastically if the information of the supporting graph topology is taken into account. 
\end{abstract}
\section{Introduction}\label{S:intro}

    Ours is a digital and pervasively connected world where an unprecedented amount of information is generated, recorded and processed. Coping with the challenges associated with complex contemporary datasets calls for new models and tools able to, among other things, account for irregular information domains and nonlinear interactions. Motivated by these goals, this paper leverages recent successes in the fields of graph signal processing (GSP) and deep learning (DL) to propose a new architecture for (underparametrized) deep decoding of signals supported on graphs.     
    
    The main goal of GSP is to generalize classical SP results to signals defined on irregular domains \cite{shuman2012theemergingfield,djuric2018bookGSP}. The underlying assumption in GSP is twofold: i) the geometry of the irregular domain can be well represented by a graph (i.e., by local pairwise interactions) and ii) the properties of the signal of interest depend on the topology of such a graph.  A plethora of graph signals exist, with examples ranging from neurological activity patterns defined on top of brain networks to the spread of epidemics over social networks \cite{shuman2012theemergingfield}. Assuming complete knowledge of the graph, early GSP efforts focused on the rigorous definition and analysis of linear graph filters, graph Fourier representations, and their application to inverse problems such as denoising, deconvolution, or sampling and reconstruction, to name a few \cite{SandryMouraSPG_TSP14Freq,chen2015discrete, segarra2017filtering, segarra2017blind}. More recent efforts focus on identifying the supporting graph, developing robust GSP schemes, analyzing higher-dimensional graph signals, and developing nonlinear GSP architectures. Tractable and useful examples of the latter category are median filters \cite{segarra2016center-weighted} and DL schemes \cite{bronstein2017geometricdeeplearning}, which have been mostly focused on (graph-filter based) convolutional architectures \cite{defferrard2016convolutional,gama2019convolutionalgraphs}, see, e.g., \cite{tian2014learning,kipf2016variational} for some exceptions. The interest in DL architectures is not surprising since in the last years, propelled by advances in data availability, computation, and optimization, deep neural network architectures have emerged as one of the most powerful tools for machine learning over modern datasets \cite{lecun2015deep}. Early works addressed classification and regression tasks, especially for speech and images, where convolutional architectures have shown a remarkable performance \cite{lecun2015deep,goodfellow2016deep}. Subsequent works went beyond supervised tasks, including representation learning, reinforcement learning and development of deep latent models. Two notable examples within this later category are generative (adversarial) networks and deep autoencoders \cite{goodfellow2014generative,doersch2016tutorial,kingma2013auto}, which are extremely useful to address inverse problems where a low-dimensional representation of the signal of interest is required \cite{hinton2006reducing}. Generally, most of the success of DL for image and speech processing is attributed to the high availability of data, which enables the training of overparametrized architectures.
   However, it was recently shown that the structure of convolutional~\cite{ulyanov2018deep} and non-convolutional~\cite{heckel2018deep} decoder (generator) networks can effectively capture many low-level image features even prior to any training. 
   
   Inspired by these lines of work, our main contribution is the design of the first underparametrized deep decoder architecture for graph signals. By leveraging agglomerative hierarchical clustering methods, the proposed architectures provide a topology-aware nonlinear representation basis for graph signals that is more efficient than the graph Fourier transform (GFT) and than can be used for tasks such as \textit{compression}, \textit{denoising}, and inpainting of signals on graphs.

\section{Preliminaries}\label{S:preliminaries}
This section introduces notation and reviews basic GSP and deep decoding concepts that will be used in the ensuing sections. 
\subsection{Graph Signal Processing}\label{Ss:GSP}

Let $\ccalG=(\ccalN,\ccalE)$ denote a directed graph, where $\ccalN$ is the set of nodes, with cardinality $N$, and $\ccalE$ is the set of edges, with $(i,j)\in\ccalE$ if $i$ is connected to node $j$. The set $\ccalN_i:=\{\, j \, |(j,i)\in\ccalE\}$ denotes the incoming neighborhood of node $i$. For a given $\ccalG$, the adjacency matrix $\bbA \in \reals ^{N\times N}$ is sparse with non-zero elements $A_{ij}$ if and only if $(j,i)\in\ccalE$. If $\ccalG$ is unweighted, the elements $A_{ij}$ are binary. If the graph is weighted, then the value of $A_{ij}$ captures the strength of the link from $j$ to $i$. The focus of this paper is not on analyzing $\ccalG$, but a graph signal defined on its set of nodes. Such a signal can be represented as a vector $\bbx=[x_1,\ldots,x_N]^T \in  \mathbb{R}^N$ where the $i$-th entry represents the signal value at node $i$. Since the signal $\bbx$ is defined on the graph $\ccalG$, the underlying assumption in GSP is that the properties of $\bbx$ depend on the topology of $\ccalG$. For instance, if the graph encodes similarity and the value of $A_{ij}$ is high, then one would expect the signal values $x_i$ and $x_j$ to be closely related. 


\vspace{.1cm}\noindent \textbf{The graph-shift operator (GSO).}
The GSO $\bbS$ is defined as an $N\times N$ matrix whose entry $S_{ij}$ can be non-zero only if $i=j$ or $(j,i)\in\ccalE$. 
Common choices for $\bbS$ are $\bbA$ and the graph Laplacian $\bbL$, which is defined as  $\bbL:=\diag(\bbA\bbone)-\bbA$ \cite{shuman2012theemergingfield,djuric2018bookGSP}. The GSO accounts for the topology of the graph and, at the same time, represents a linear transformation that can be computed \textit{locally}. Specifically, if $\bby=[y_1,\ldots,y_N]^T$ is defined as $\bby=\bbS\bbx$, then node $i$ can compute $y_i$ provided that it has access to the values of $x_j$ at its neighbors $j\in \ccalN_i$. We assume that $\bbS$ is diagonalizable so that there exists an $N\times N$ matrix $\bbV$ and a diagonal matrix $\bbLambda$ such that $\bbS=\bbV\bbLambda\bbV^{-1}$.


\vspace{.1cm}
\noindent\textbf{Frequency domain representation.} Key to define the frequency representation of graph signals and graph filters is the eigendecomposition of $\bbS$.
In particular, 
the frequency representation of a graph \emph{signal} $\bbx$ is defined as $\tilde{\bbx}:=\bbV^{-1}\bbx$
, with $\bbV^{-1}$ 
acting as the GFT~\cite{SandryMouraSPG_TSP14Freq}. 
%
%
Leveraging the GFT, one can the generalize the notion of bandlimitedness to the graph domain. Specifically, it is said that $\bbx$ is a low-pass bandlimited graph signal if it holds that its frequency representation $\tbx=\bbV^{-1}\bbx$ satisfies that $\tdx_k=0$ for all $k>K$, where $K\leq N$ represents the bandwidth of the signal $\bbx$. Note that if $\bbx$ is bandlimited with bandwidth $K$, then it holds that 
\begin{equation}\label{E:bandlimited_graphsignals}
\bbx=\bbV_K\tbx_K,
\end{equation} 
where $\tbx_K:=[\tdx_1,...,\tdx_K]$ collects the active frequency components and the $N\times K$ matrix $\bbV_K:=[\bbv_1,...,\bbv_K]$ collects the corresponding eigenvectors. In other words,  \eqref{E:bandlimited_graphsignals} states that: i) the original $N$-dimensional signal $\bbx$ lives on a (reduced-dimensionality) subspace related to the spectrum of the supporting graph; and ii) the $K$ values in $\tbx_K$ suffice to represent $\bbx$. This reduced-dimensionality representation has been exploited to design efficient algorithms for denoising, reconstruction from a subset of samples, and other inverse problems dealing with graph signals \cite{marques2016sampling,chen2015discrete}. While the simple linear model in \eqref{E:bandlimited_graphsignals} has been shown relevant in real-world applications \cite{djuric2018bookGSP}, many modern datasets have a \emph{nonlinear} latent structure that linear reduced-dimensionality models such as the one in \eqref{E:bandlimited_graphsignals} are not able to capture.

\subsection{Deep Decoding}\label{Ss:DeepDec}

Deep decoders are capable of learning nonlinear reduced dimensionality representations to approximate a dataset (collection of signals) of interest. Succinctly, the idea behind deep decoding is twofold: i) the signals of interest can be represented by a vector of latent variables whose dimension is much smaller than that of the original signals and ii) the mapping from the latent space to the observable signal space is given by a succession of layers, each of them consisting of a (learnable) linear transformation followed by a (pre-specified) scalar point-wise nonlinearity. This reduced representation can be later exploited to compress the signals of interest or, alternatively, to solve underdetermined inverse problems, which, after assuming such parsimonious representation, typically become well posed.  

To be mathematically rigorous, let $\bbx\in\reals^N$ represent the signal of interest and let $\bbZ \in \reals^{N_0\times F_0}$ be a matrix containing the hidden variables with $(N_0F_0)<< N$. The assumption is therefore that the signal $\bbx$ can be written (or well approximated) as
\begin{equation}
\bbx = f_{\bbTheta}(\bbZ),    
\end{equation} 
where $f_{\bbTheta}(\cdot)$ is a \emph{parametric nonlinear} function whose parameters are collected in  $\bbTheta$. For the particular case of deep decoders, and with $L$ denoting the total number of layers, the function $f_{\bbTheta}: \reals^{N_0\times F_0} \rightarrow \reals^N$ is assumed to be given as
\begin{align}
\bbY^{(0)}&=\bbZ,\label{E:DeepDecoderNonGraph_input}\\
\hbY^{(l)}&= \ccalT_{\bbphi^{(l)}}^{(l)} \Big\{\bbY^{(l-1)}\Big\},\;\;1\leq l \leq L,\\
\check{\bbY}^{(l)}_{ij} &= g^{(l)} \Big([\hbY^{(l)}]_{ij}\Big),\;\;1\leq l \leq L\; \text{and}\; \text{all}\;ij,\\
\bbY^{(l)} &= \ccalF_{\bbpsi^{(l)}}\Big\{\check{\bbY}^{(l)}\Big\},\;\;1\leq l \leq L,\;\\
\label{E:DeepDecoderNonGraph_output} \bbx &= \bbY^{(L)},\; 
\end{align}
where $\bbZ$ is the input, $\bbx$ is the output, $l$ is the layer index, $\ccalT_{\bbphi^{(l)}}^{(l)}\{\cdot\}: \reals^{N_{l-1}\times F_{l-1}} \to \reals^{N_{l}\times F_{l}}$ is the linear transformation implemented at layer $l$, $\bbphi^{(l)}$ are the parameters that define such a transformation, $g^{(l)}: \reals \rightarrow \reals$ is a scalar nonlinear operator (possibly different per layer), and $\ccalF_{\bbpsi^{(l)}}\{\cdot\}$ is a simple function that re-scales and modifies the mean of each column of $\check{\bbY}^{(l)}$ using parameters $\bbpsi^{(l)}\in\reals^{F_l\times2}$. As a side note, such a function is oftentimes not implemented in the last layer. Hence, if $\bbTheta^{(l)}$ collects the values of $\bbphi^{(l)}$ and $\bbpsi^{(l)}$ and $\bbTheta$ is defined as $\bbTheta:=\{\bbTheta^{(l)}\}_{l=1}^L$, the output $\bbx$ generated by the deep decoder function $f_{\bbTheta}$ when the input is $\bbZ$ can be effectively computed using \eqref{E:DeepDecoderNonGraph_input}-\eqref{E:DeepDecoderNonGraph_output}.   

\section{Proposed architecture}
\label{S:joint_estimation}

\begin{figure}[]
    \centering
    \includegraphics[width=0.4\textwidth]{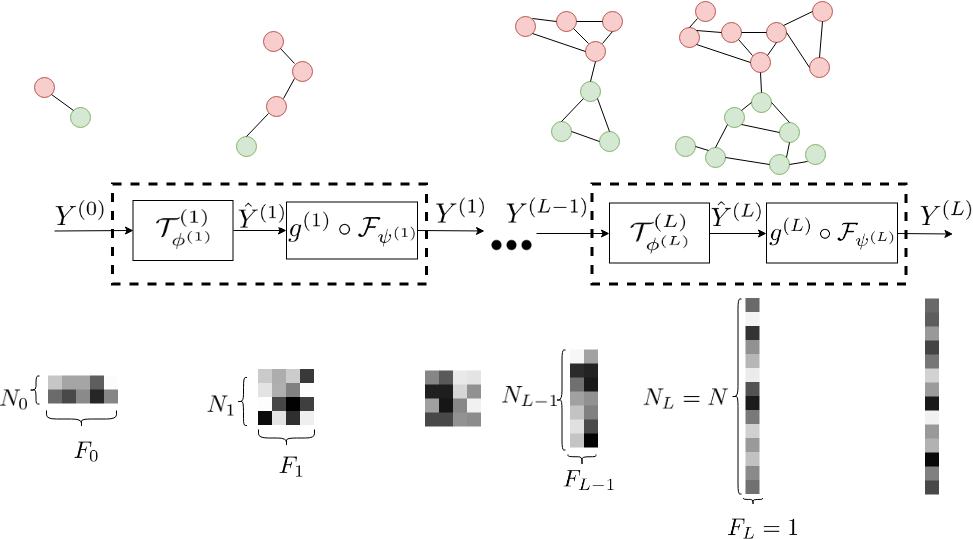}
    \vspace{-.25cm}
	\caption{Scheme representing the proposed deep decoding architecture for graph signals. The top part of the figure represents the support of the signals. The middle part represents the transformations involved at each of the layers. The bottom part depicts the values and size of the signals. }
	\label{fig:pictorialrepresentation_DD_architecture_GS}
	\vspace{-.25cm}

\end{figure}

The main contribution of this paper is the development of an \textit{underparametrized} deep decoding architecture \textit{for signals supported on a graph} $\ccalG$. Inspired by \cite{heckel2018deep}, which deals with representation of images defined in a 2-D grid, we propose a deep decoding architecture with $L$ layers (see Figure \ref{fig:pictorialrepresentation_DD_architecture_GS} for a pictorial representation) where:
\begin{itemize}
\item[i)] The input $\bbZ\in \reals^{N_0\times F_0}$ is generated as a random realization of a process that is white across rows and columns
\item[ii)] The linear operator at layer $l$ is defined as
\begin{equation}
\ccalT_{\bbphi^{(l)}}^{(l)} \{\bbY^{(l-1)}\} = \bbU^{(l)} \bbY^{(l-1)} {\bbPhi^{(l)}}^T. \label{E:factorizedlinearoperatordeepdec_graphs}
\end{equation}
\end{itemize}
In the GSP setup considered in this paper, the input feature matrix $\bbY^{(l-1)}\in \reals^{N_{l-1}\times F_{l-1}}$ in \eqref{E:factorizedlinearoperatordeepdec_graphs} represents $F_{l-1}$ graph signals, each of them defined on a graph $\ccalG_{l-1}$ with $N_{l-1}$ nodes. Similarly, the output matrix $\hbY^{(l)}= \ccalT_{\bbphi^{(l)}}^{(l)} \{\bbY^{(l-1)}\}\in \reals^{N_{l}\times F_{l}}$ contains $F_{l}$ graph signals, each of them defined on a graph $\ccalG_{l}$ with $N_{l}$ nodes. Regarding the particular form of the linear transformation $\ccalT_{\bbphi^{(l)}}^{(l)}$, we note that the \eqref{E:factorizedlinearoperatordeepdec_graphs}  postulates a factorized (separated) operation across rows (nodes) and columns (graph signals). In particular, we note that $\bbPhi^{(l)}\in \reals ^{F_{l} \times F_{l-1}}$ represents a mixing fat matrix that generates new $F_{l}$ features  by combining linearly the $F_{l-1}$ features of the previous layer.  More importantly for the setup at hand, the tall matrix $\bbU^{(l)} \in \reals ^{N_{l} \times N_{l-1}}$ represents an upsampling operator that maps the values of the graph signal at layer $l-1$ to the graph signal at layer $l$.
The design of $\{\bbU^{(l)}\}_{l=1}^L$ the subject of the next section.

\subsection{Nested collection of upsampling graph-signal operators}\label{Ss:collection_upsampl_GS_operators}

\begin{figure}[]
    \centering
    \includegraphics[width=0.3\textwidth]{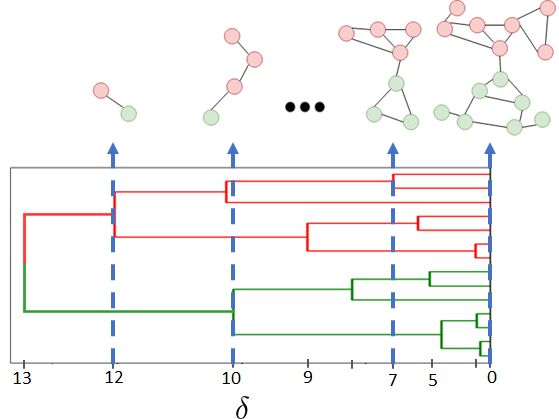}
    \vspace{-.2cm}
	\caption{Dendrogram of a hierarchical agglomerative clustering algorithm and the resulting graphs with 14, 7, 4 and 2 nodes.}
	\label{fig:dendrogram_with_graphs}
\end{figure}

While in regular grids the upsampling operator is straightforward, when the signals at hand are defined on irregular domains the problem becomes substantially more challenging.
To generate topology-aware upsampling operators, we will rely on agglomerative hierarchical clustering methods~\cite{jain_1988_algorithms, carlsson_2018_hierarchical}.
These methods take as input a graph and output a dendrogram, which can be represented as a rooted tree; see Fig.~\ref{fig:dendrogram_with_graphs}.
The interpretation of a dendrogram is that of a structure which yields different clusterings at different resolutions. 
At resolution $\delta = 0$ each point (node) is in a cluster of its own. As the resolution
parameter $\delta$ increases, nodes start forming clusters. Eventually, the resolutions become coarse enough so that all nodes become members of the same cluster. 
By cutting the dendrogram at $L+1$ resolutions (including $\delta = 0$) we obtain a collection of node sets with parent-child relationships inherited by the refinement of clusters. E.g., the three red nodes in the second graph are the children of the red parent in the coarsest graph in Fig.~\ref{fig:dendrogram_with_graphs}.
From these relationships, we define the matrices $\bbP^{(l)} \in \{0,1\}^{N_{l} \times N_{l-1}}$ where the entry $P^{(l)}_{ij}$ is $1$ if node $i$ in layer $l$ is the child of node $j$ in layer $l-1$, and $0$ otherwise.
Upon defining a (row-normalized) adjacency matrix $\bbA^{(l)}$ between the clusters at layer $l$, the upsampling operator is given by
\begin{equation}\label{E:upsampling_operator}
\bbU^{(l)} = (\gamma \bbI + (1-\gamma) \bbA^{(l)}) \bbP^{(l)},
\end{equation}
where $\gamma$ is a pre-specified constant. Notice that $\bbU^{(l)}$ in \eqref{E:upsampling_operator} first copies the signal value of parents to children ($\bbP^{(l)}$) and then every child performs a convex combination between this value and the average signal value of its neighbors. In our numerical experiments, we will test three different ways of defining $\bbA^{(l)}$ for intermediate layers: i) `NoA', where $\bbA^{(l)} = \bbI$, ii) `Bin', where two nodes have an unweighted edge if \emph{any} of their corresponding (grand)children in the original graph ($\delta=0$) have an edge between them, and iii) `Wei', which is a weighted variant of `Bin' depending on the number of edges between (grand)children.

\subsection{Deep coding of a graph signal }
Suppose now that we are given a signal $\bbx\in\reals^N$ defined on graph $\ccalG$ and our goal is to encode $\bbx$ using the architecture defined in \eqref{E:DeepDecoderNonGraph_input}-\eqref{E:factorizedlinearoperatordeepdec_graphs} relying on the method to design the upsampling matrices $\{\bbU^{(l)}\}_{l=1}^L$ proposed in Section \ref{Ss:collection_upsampl_GS_operators}. Alg. \ref{alg:graph_deep_decoder} describes the steps to carry out that task. Clearly, Alg. \ref{alg:graph_deep_decoder} can be used for compression, but also to handle inverse problems such as denoising
, as described next.

\begin{algorithm}[t]
\SetKwInOut{Input}{Input}
\SetKwInOut{Output}{Output}
\Input{$\ccalG$ and $\bbx$.}
\Output{$\hbx$ and $\{\hbTheta^{(l)}\}_{l=1}^L$.}
\SetAlgoLined
 Use $\ccalG$ to generate $\{\bbU^{(l)}\}_{l=1}^L$ according to~\eqref{E:upsampling_operator}\\
 Generate random $\bbZ $ independently across rows and columns.\\
 Randomly initialize $\{\hbTheta^{(l)}_0\}_{l=1}^L$ \\
 Select desired loss function $\ccalL$ \\
 \For{$t=0$ \KwTo $T^{\max}$}{
 Update $\{\hbTheta^{(l)}_{t+1}\}_{l=1}^L$ by minimizing $\ccalL(\bbx, f_{\hbtheta_i}(\bbZ))$ with an stochastic gradient descent optimizer.
}
$\hbx = f_{\hbtheta_{T^{\max}}}(\bbZ)$ \\
$\{\hbTheta^{(l)}\}_{l=1}^L = \{\hbTheta^{(l)}_{T^{\max}}\}_{l=1}^L$
\caption{Short description of the Graph Deep Decoder.}
\label{alg:graph_deep_decoder}
\end{algorithm}

\begin{itemize}[leftmargin=*]
\item
If compressing $\bbx$ is the ultimate goal, the loss function can be simply set to $\ccalL(\bbx,\bbx')\!=\!\|\bbx - \bbx'\|_2^2$ and the matrices $\{\bbTheta^{(l)}\}_{l=1}^L$ obtained after running Alg. \ref{alg:graph_deep_decoder} are the output of interest, with the resulting compression ratio being $\eta\!=\!(\sum_{l=1}^L F_{l}(F_{l-1}+2))/N$.

\item Suppose now that the observed signal is $\bbx=\bbx_o + \bbw$ with $\bbx_o\in\reals^N$ being the signal of interest and $\bbw\in\reals^N$ being an additive noise of known covariance $\bbR_{ww}$. The goal in this setup is to use the proposed architecture to denoise $\bbx$, so that a better estimate of $\bbx_o$ is obtained. In this case, it suffices to run Alg. \ref{alg:graph_deep_decoder} with $\bbx=\bbx_o + \bbw$ as input, use the generated output $\hbx$ as estimate of $\bbx_o$, and set the loss function to $\ccalL(\bbx,\bbx') = \|\bbR_{ww}^{-1/2}(\bbx-\bbx')\|_2^2$.

\end{itemize}

\begin{figure*}[t]
	\centering
	\begin{subfigure}{0.32\textwidth}
		\centering
		    \includegraphics[width=1\textwidth]{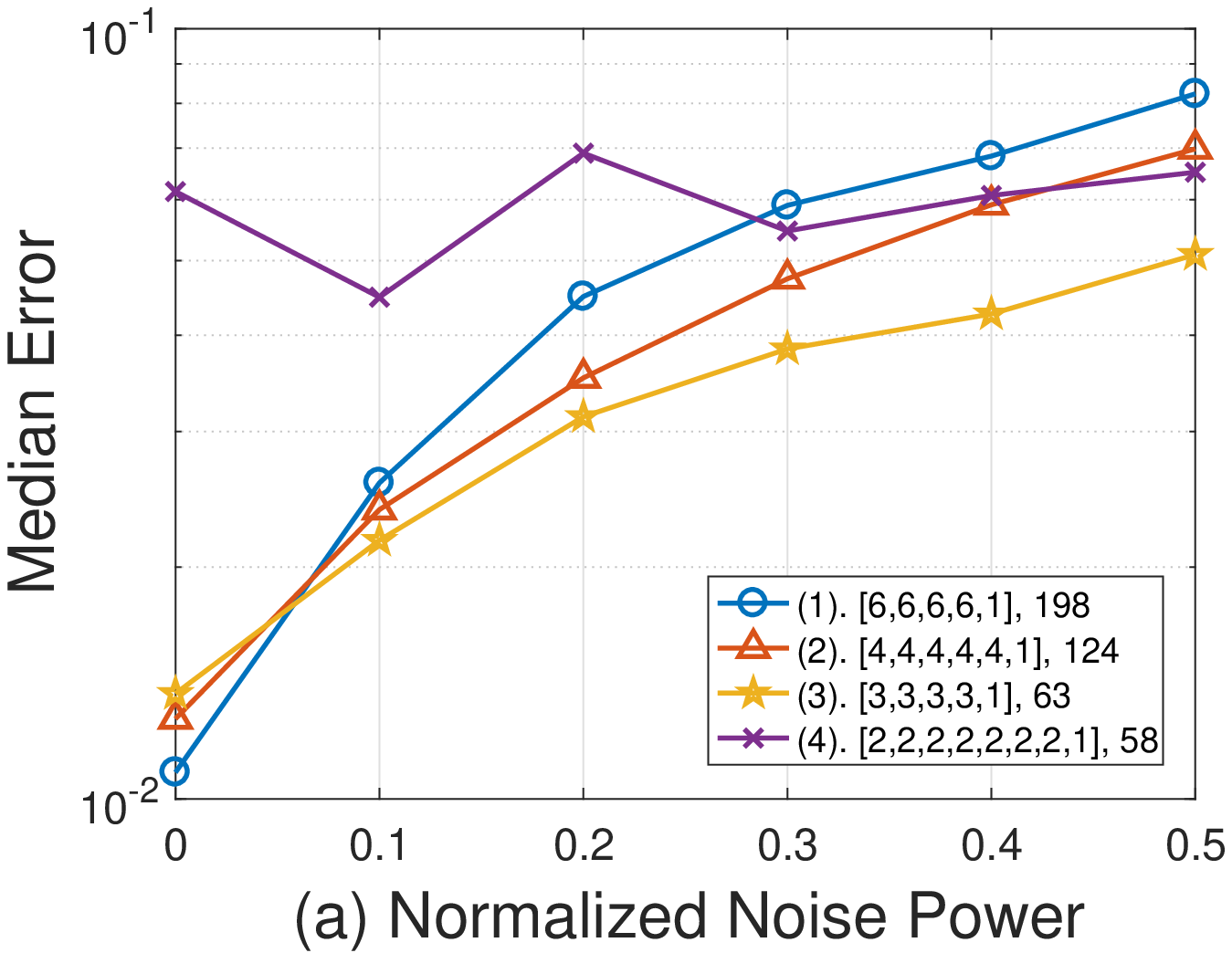}
	\end{subfigure}
	\begin{subfigure}{0.32\textwidth}
		\centering
			\includegraphics[width=1\textwidth]{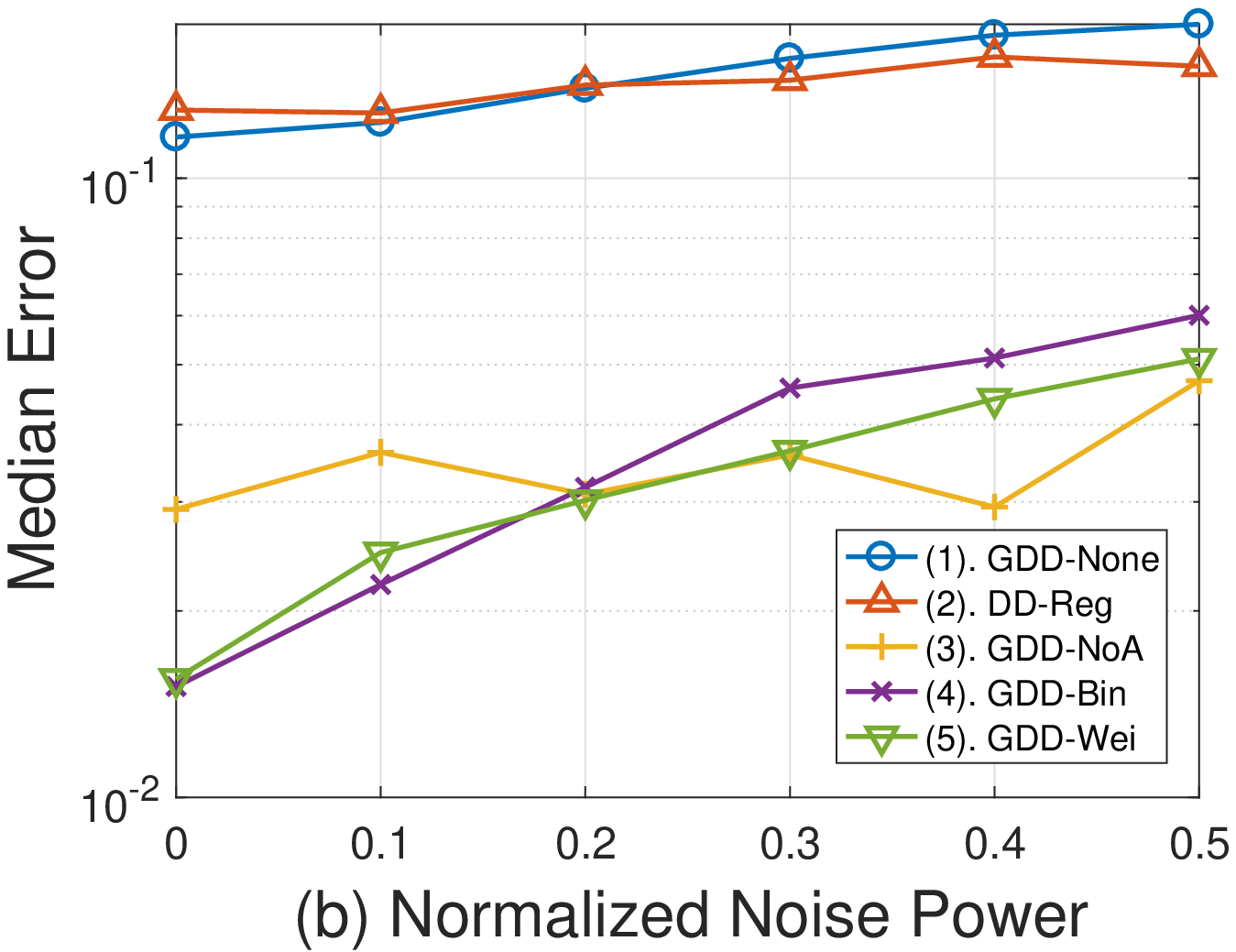}
	\end{subfigure}
	\begin{subfigure}{0.32\textwidth}
		\centering
			\includegraphics[width=1\textwidth]{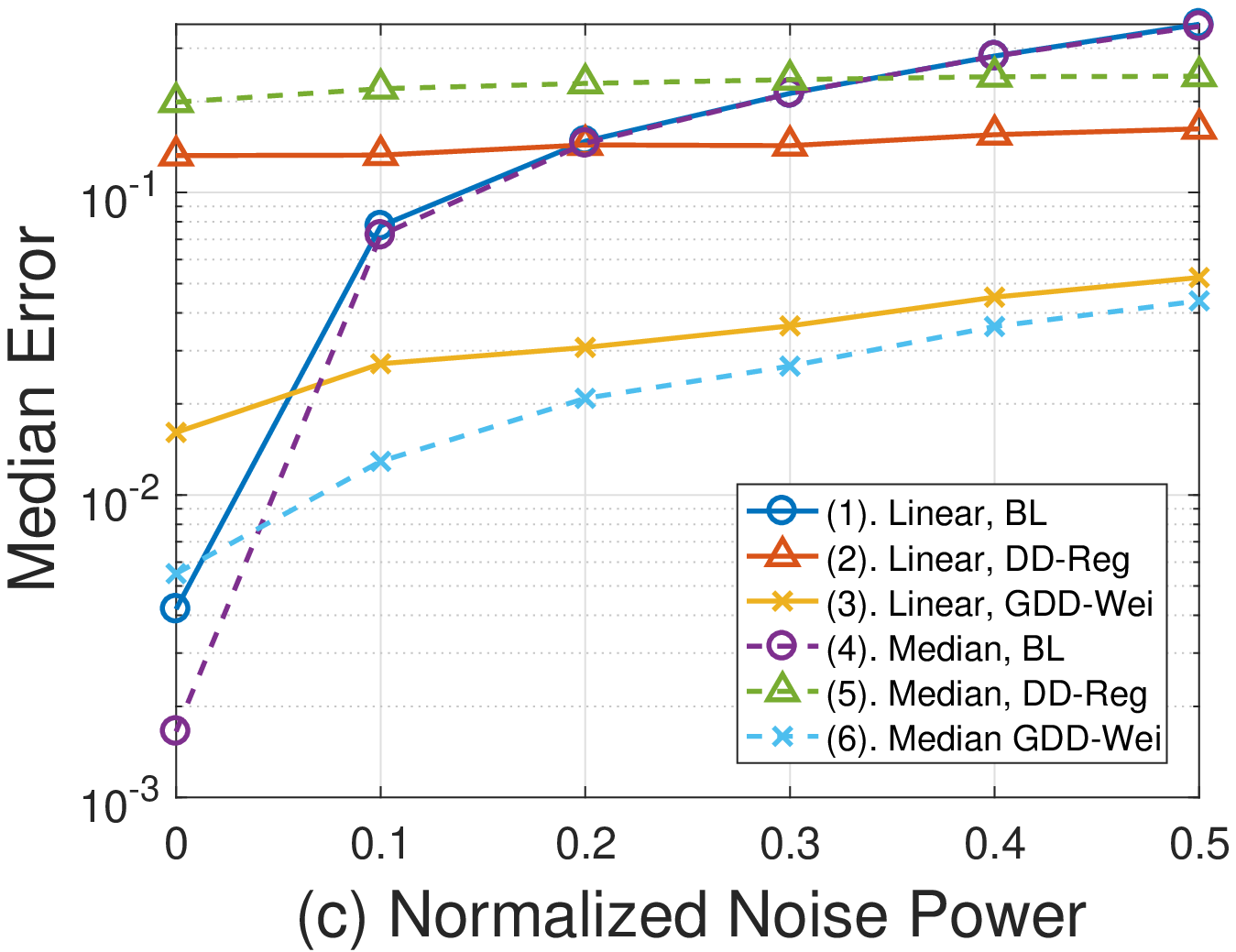}
	\end{subfigure}
	\vspace{-.25cm}
	\caption{Error after denoising for synthetic data and different levels of noise. One hundred signals defined over an SBM graph with $N=256$ nodes are tested. 
	(a) Impact of the complexity of the architectures using a weighted upsampling. The legend specifies the number of features per layer $[F_1,...,F_L]$ and the total number of trainable parameters.
	(b) Impact of the upsampling method using an architecture with 63 parameters. (c) Impact of the type of signal using an architecture with 63 parameters and a weighted upsampling.}
	\label{fig:synt_exps}
	\vspace{-.35cm}
\end{figure*}

\begin{figure}[]
    \centering
    \includegraphics[width=0.32\textwidth]{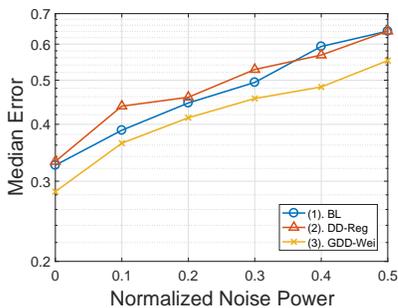}
    \vspace{-.2cm}
	\caption{Error after denoising for real-world data from D\&D dataset \cite{dobson2003distinguishing}. Only connected graph with at least 50 nodes are considered. One hundred signals with a medium size of 107 have been tested.}
	\label{fig:real_data}
	\vspace{-.37cm}
\end{figure}
\section{Numerical experiments}\label{S:num_experiments}
In this section we present experiments to evaluate the performance of the proposed underparametrized deep encoder, showcasing the benefits of including the information of the underlying graph. The code to reproduce the experiments is available on GitHub\footnote{\url{https://github.com/reysam93/Graph_Deep_Decoder}}.

We start by characterizing our architecture using synthetic data. The obtained results are summarized in the 3 panels of Fig. \ref{fig:synt_exps}. In all three settings, we considered a stochastic block model (SBM) graph with 256 nodes and 4 communities \cite{decelle11}. Edges exist with probability 0.15 if the incident nodes are in the same community and with probability $2.25\times 10^{-3}$ otherwise. The recovery error associated with an observed noisy signal $\bbx=\bbx_o + \bbw$ is evaluated as $\|\hbx - \bbx_o\|_2^2$, with the value reported in the plots being the median error across $100$ generated signals.

Fig. 3(a) plots the denoising performance of 4 architectures with different complexity (i.e., different number of layers and trainable parameters). First and foremost, the results validate our graph-aware approach demonstrating that, in the absence of noise, the approximation error is in all cases below 0.05 ($95\%$ accuracy), even for compression ratios $\eta$ in the order of $25\%$. Moreover, we observe that the most complex model has the smallest error in the absence of noise, but its performance deteriorates quickly as the noise power increases. On the other hand, the simpler architectures are more robust to noise, despite having a larger noiseless error due to their higher compression loss. The architecture's trade off is then clear: the number of parameters must be large enough to learn a good compressed representation of the signal, but small enough so that the noise is not learnt. With this in mind, the third architecture is specially interesting: it achieves a compression ratio of $25\%$, attains one of the smallest compression errors and is robust to noise. Hence, this will be the architecture used for the remaining simulations.

Fig. 3(b) evaluates the influence of the upsampling operator. The options tested are: (i) `GDD-None': a graph deep decoder model as in \eqref{E:factorizedlinearoperatordeepdec_graphs} but without upsampling (i.e.., $\bbU^{(l)}=\bbI$ for all $l$); (ii) `DD-Reg': a deep decoder model with linear upsampling
proposed in \cite{heckel2018deep}; (iii) `GDD-NoA'; (iv) `GDD-Bin'; and (v) `GDD-Wei'. The tree last options are graph deep decoder (GDD) schemes where $\bbU^{(l)}$ are constructed as in \eqref{E:upsampling_operator}, with $\gamma=0.5$ and $\bbA^{(l)}$ being defined using the methods `NoA', `Bin' and `Wei' presented in Sec. \ref{Ss:collection_upsampl_GS_operators}.
It can be seen clearly how methods (i) and (ii), which ignore the structure of the underlying graph, exhibit the worst performance. We also note that, for low noise levels, the error of method (iii) is larger than that of methods (iv) and (v). This is not surprising because the upsampling in (iii) incorporates less information about $\ccalG$ than those in (iv) and (v). In addition, to better asses the stability of the results, we computed the relative distances between error percentiles as $ \frac{pctl75 - pctl25}{median}$. The results for (iii), (iv) and (v) are \textit{0.65}, \textit{0.4} and \textit{0.39}, respectively, showing that the third method is also more unstable. Finally, because the `Wei' upsampling slightly outperforms the `Bin' upsampling, the former is selected for the remaining tests. 

Fig. 3(c) assesses the reconstruction performance for different types of signals. `Linear' signals are created following a linear diffusion process of the form $\bbx_{Linear}=\sum^{T-1}_{t=0}h_t\bbA^t\bbs$, where $\bbs$ is a sparse signal and $T=6$ \cite{segarra2017blind}. `Median' signals are created from linear signals where the value of each node is the median of its neighbours. In addition, we compare our scheme with a deep decoder with regular upsampling and with a linear graph-aware bandlimited reconstruction as in \eqref{E:bandlimited_graphsignals}. In the absence of noise the error of the bandlimited model is smaller, and hence, we can assume that both signals are bandlimited. However, when noise is added, the graph deep decoder obtains the best results. It can also be observed that for deep regular decoders the nonlinear transformation entails an increment on the error, while for our architecture it has the opposite effect. This can be explained because the values of the `Median' signals are even more related to the graph topology and, therefore, incorporating the graph information is more relevant. 

We close this section by presenting an experiment using the graph and signals from the D\&D protein structure database \cite{dobson2003distinguishing}, where nodes represent amino acids, links capture their similarity, and signals are their expression level. Using this database, we compare the performance of: (i) a bandlimited recontruction; (ii) an underparametrized deep decoder reconstruction; (iii) an underparametrized graph deep decoder with weighted upsampling reconstruction. Because the minimum size of the graphs is of 50 nodes, an architecture with 48 parameters have been used, instead of the one presented in the previous simulations, achieving an average compression rate of 0.5. The results are shown in Fig. \ref{fig:real_data}, illustrating that the architecture proposed in this paper outperforms the considered alternatives, especially for low noise values.
 

\vfill\pagebreak
\balance
\bibliographystyle{IEEEbib}
\bibliography{CAMSAP_graph_decoder}

\end{document}